\documentclass{desyproc}

\begin{document}
\title{Theory of Dark Matter$~$\footnote{Plenary review talk  given at the  ``Physics at  the LHC 2010" conference,
7-12 June 2010, DESY, Hamburg.}}

\author{{\slshape Graciela B Gelmini}  \\[1ex]
 Department of Physics and Astronomy,
University of California Los Angeles (UCLA),
 475 Portola Plaza,
 Los Angeles, CA 90095, USA }

\acronym{PLHC2010}

\maketitle

\begin{abstract}
The search for dark matter is a very wide and active field of research. Many  potential hints of dark matter have appeared recently which led to a burst of theoretical activity and model building.  I necessarily concentrate here only in some aspects of it. I review here some recent hints  and some of the ways in which they could be explained.
\end{abstract}

\section{Elements of a theory of dark matter}

We know a lot about dark matter (DM) but we  still do not  have a clue of what it consists of.  We know the abundance of the DM in the Universe at the level of  few percent, 
  $\Omega_{\rm CDM} = 0.232 \pm 0.013$~\cite{Komatsu2010} and that  it is not baryonic.
We know also that the  DM cannot be explained within the Standard Model (SM) of elementary particles. 
The bulk of the DM can only be either Cold (CDM) or Warm (WDM), namely it was non-relativistic or becoming non-relativistic at the moment galaxies should start forming in the early Universe, at temperatures $T \simeq$ keV. 
There are no WDM or CDM candidates in the SM, but there are many in all extensions of the SM. For example, sterile neutrinos and gravitinos  can be WDM. WIMPs, Weakly Interacting Massive Particles, among others,
could be CDM.  

The argument showing that WIMPs  are good DM candidates is old and well known.  The density per comoving volume of non-relativistic  particles in equilibrium in the early Universe decreases exponentially with decreasing temperature, due to the Bolzmann factor, until the reactions which change the particle number become ineffective.   At this point,   
 the WIMP number per comoving volume becomes constant. This  moment of chemical decoupling or freeze-out happens later, i.e. for smaller WIMP densities, for larger annihilation cross sections $\sigma$
  and the present (standard) relic density is 
$\Omega_{\rm std}  \simeq 0.2 \times  [(3 \times 10^{-26} {\rm ~cm}^3/{\rm s})/ \left< \sigma v \right>] $, which for weak cross sections $\sigma \simeq 3 \times 10^{-26} {\rm cm}^3/{\rm s}$ gives the right 
DM density (and a temperature  $T_{f.o.} \simeq m_\chi/20$ at freeze-out for a WIMP of mass $m_\chi$). Some call this ``the WIMP miracle". However, the WIMP relic density depends not only on the particle model 
but on the history of the Universe before Big Bang Nucleosynthesis (BBN), an epoch from which we have no data. BBN is the earliest episode (finishes 200 s  after the Bang, when  $T\simeq 0.8$ MeV) from which we have a trace, the abundance of light elements. WIMP's have their number fixed at $T_{f.o.}$, thus WIMPs with $m_\chi \geq 100$ MeV would be the earliest remnants and, if discovered,  they would  for the first time give  information on the pre-BBN epoch of the Universe. At present, to compute the WIMP relic density we must make assumptions about the pre-BBN epoch. The standard relic density is derived  assuming that  the entropy of matter and radiation is conserved,
that WIMPs are produced thermally, i.e. via interactions with the particles in the plasma,  and were in kinetic and chemical equilibrium before they decoupled at $T_{f.o.}$. WIMPs produced in this way are called ``thermal WIMPs". The standard assumptions do not hold in many viable pre-BBN cosmological models, and in some of those, WIMPs can have very different relic densities (e.g. neutralinos can have the DM density in practically all supersymmetric  models~\cite{ggsy}).

Because of spontaneous symmetry breaking arguments totally independent of the DM issue, we do expect new physics beyond the SM to appear at the electroweak scale soon to be explored by the Large Hadron Collider (LHC) at CERN.  Naturalness arguments imply that above the TeV scale there should be cancellations in the radiative corrections to the SM Higgs mass due to a new theory, such as supersymmetry, 
 technicolor,  large extra spatial dimensions or the Little Higgs model, for example.
  These extensions of the SM provide the main potential discoveries at the LHC and also DM candidates that are sometimes described as ``well motivated". However, as shown in many of the recent models exclusively motivated by DM hints, the new physics needed to explain the DM may be very different. We will in the following take as a paradigm of this new physics the model ambitiously named precisely  ``A Theory of DM"~\cite{Arkani-Hamed}. 
 This  model  is made to fit DM data, not to solve the SM hierarchy problem. Besides it provides signatures for the LHC, which depend on the particular realization of the model~\cite{Arkani-2}. Thus, physics beyond the SM is required by the DM and expected at the electroweak scale, but both new physics may or may not be related. The experiments at the LHC and the searches for the DM in our galactic halo are independent and complementary.

 Direct DM searches look for energy deposited within a detector by the collisions  of  WIMPs belonging to the dark halo of our galaxy.  I will mention the  DAMA modulation signal, the possible hint seen by CoGeNT and bounds from Xenon10, Xenon100 and CDMS~\cite{Baudis}.  Indirect DM searches look for WIMP annihilation (or decay) products.
  I will concentrate here on the  positron data of PAMELA and Fermi and models to explain them.

\section{Dark matter hints from direct searches}

The DAMA collaboration, in the 13 years of combined data of  the DAMA/NaI and DAMA/Libra experiments,
 has found a 8.9$\sigma$ annual modulation signal~\cite{Bernabei:2010}
 compatible with the signal expected from DM particles bound to our
galactic halo and a standard halo model (due to the motion of the Earth around the Sun). 
Are the DAMA results compatible with those of all other searches? There are many aspects to this question and  I will concentrate on two possibilities: inelastically scattering DM (IDM),  and light  elastically scattering WIMPs. 

In  IDM models~\cite{IDM}, in addition to the DM state $\chi$ with mass $m_\chi$  there is an excited state $\chi^*$, with a small mass difference $m_\chi^* -m_{\chi}=\delta \simeq 100~{\rm keV}$ and the 
inelastic scattering with a nucleon $N$, $\chi +N \to \chi^* +N$, dominates over  the elastic scattering.
While the minimum WIMP velocity necessary to provide a particular recoil energy $E_R$ in an elastic collision is $v^{el}_{min}=  \sqrt{ ME_R/2\mu^2} $, the minimum WIMP velocity required for an inelastic collision is higher  $v^{inel}_{min}=  v^{el}_{min}  + \delta/\sqrt{2M E_R}$. Thus, only high-velocity DM particles have enough energy to up-scatter. $v^{inel}_{min}$ grows as $E_R$ decreases, so there are no low $E_R$ events and the spectrum is very different
than for elastic collisions. Besides, $v^{inel}_{min}$ decreases with increasing target mass $M$, thus targets with high mass are favored (better I in DAMA than Ge in CDMS).
 The  modulation of the signal is also enhanced (because the number of WIMPs changes more rapidly at high $v$), which also favors the DAMA modulation signal. For IDM with spin independent (SI) interaction with nuclei, 
a recent bound from the CDMS collaboration leaves very small room for compatibility with the DAMA signal, and new XENON100 bounds are expected soon. But there are other versions of IDM which still survive all bounds from negative searches. One example is IDM with spin dependent (SD) interactions mostly coupled to protons~\cite{IDM-SDp}. The SD coupling with a nucleus is mainly  with an unpaired nucleon, which for DAMA 
(as well as KIMS, COUPP and PICASSO)  is a p, while for XENON, ZEPLIN, CDMS and CoGeNT is a n. Thus, while inelasticity eliminates the bounds from PICASSO and COUPP, because they have light targets, the SD coupling with p eliminates those from XENON, CDMS and CRESST.

Papers written prior to the  DAMA/LIBRA results  found
regions of WIMP mass and cross section that reconciled all null
results existing at the time with  DAMA/NaI's positive signal (assuming  a standard halo
model, as is  usual  to do  to  compare experimental results). Light WIMPs with SI
interactions in the mass range 5--9~GeV~\cite{Gelmini:2004gm}, and  with SD
interactions in the mass range 5--13~GeV~\cite{Savage:2004fn} were
found to be compatible with all existing data. The situation changed
after the publication of the first DAMA/LIBRA results in 2008
(see e.g.~\cite{Savage:2008er} and reference therein). 
Light WIMPs were found incompatible with other negative results, but only at the 2 or 3$\sigma$ level, mostly when ion channeling as estimated by the DAMA collaboration~\cite{Bernabei:2007hw} was considered. A nucleus that after a collision with a WIMP recoils along the characteristic axes or planes of the crystal structure may
travel long distances without colliding with other nuclei. This channeled nucleus transfers all its energy
into electrons ($Q=1$) instead of only a fraction $Q_{Na} = 0.3$ for Na nuclei or $Q_I = 0.09$ for I nuclei, as  is the case for non-channeled recoils  ($Q$ is known as the quenching factor).  A revaluation~\cite{Bozorgnia:2010xy} of the channeling fraction in NaI has now shown that the channeling fraction is much smaller  than initially estimated by the DAMA collaboration,  thus the allowed DAMA/LIBRA region is insensitive to channeling up to the 5$\sigma$ level~\cite{Savage:2010tg}.   Besides the DAMA data, an excess of events  found recently by the CoGeNT collaboration
   (also hints in CRESST) generated renewed interest in light WIMPs,
  and a new bust of models, most having light bosons with GeV mass scale~\cite{IDM-SDp, Kuflik:2010ah}. CoGeNT is a 440g Ge detector with extremely low threshold, 
and its excess of events is  compatible with a region for WIMPs with SI interaction with mass around 10 GeV
 close to the DAMA region due to WIMP interaction with Na (with no channeling)~\cite{IDM-SDp, Kuflik:2010ah}. More data in CoGeNT, CDMS and XENON100 will clarify the situation with respect to this possible signal in the near future.

\section{Dark matter  hints from indirect searches}

The satellite INTEGRAL, launched in 2002,  has confirmed the emission of 511 keV photons from the center of the galaxy, a 30 year old signal first observed by balloon born $\gamma$-ray detectors.
 This is a clear signal of  non-relativistic positrons annihilating with electrons almost at rest.
 The isotropy of the emitting region, which initially seamed spherical and centered on the center of the galaxy, was one of the main reasons to consider DM annihilation as the origin of the positrons.
   It was argued that any astrophysical origin should show in some correlation with the visible matter distribution in the region and none had been observed until 2008, when INTEGRAL revealed an asymmetry in the emitting region, which is more extended towards the galactic plane~\cite{Weidenspointner:2008zz}.  
INTEGRAL also found evidence of a population of binary stars (called low mass X-ray binaries), known potential sources of  positrons, 
 corresponding in extent to the observed cloud of antimatter. These observations have decreased the motivation to consider DM annihilation or decay as the origin of the signal, although it is yet unclear if it can be explained satisfactorily solely with astrophysics.  Special DM candidates were proposed to explain this signal, since the annihilations of usual WIMPs would produced positrons with too high energies. Positrons must be produced with no more than a few MeV of energy. Thus, these  DM candidates either have masses of a few MeV (they are called LDM, Light DM~\cite{Boehm:2003bt}) or have an excited state   which decays to the fundamental state releasing an energy in the MeV range, although the particle is much heavier. These are called XDM, ``eXiting" DM~\cite{Finkbeiner:2007kk}).

XDM consists of a 500 GeV mass  state $\chi$  with a excited state $\chi^*$ very close in energy.
This is similar to  the ``Inelastic DM'' proposed to explain DAMA/LIBRA, but the difference in mass must be larger,
$\delta= m_{\chi^*} -m_{\chi}\sim$ MeV   and not 100 keV, so  that   $e^+ e^-$ are produced at rest  via
 de-excitation  of the excited state: $\chi^* \to \chi e^+ e^-$. The  excitation of the $\chi^*$ state is due to collisions, which fixes the particle mass, given the characteristic speed in the galaxy, $v \simeq 10^{-3} c$. Thus $\delta \simeq  (1/2) m_{\chi} 10^{-6} \simeq$ 1 MeV   which works if $m_{\chi} \simeq 500$ GeV.

Positrons and antiprotons, which would be produced in equal numbers as electrons and protons, are an interesting potential signal of WIMP annihilation because there is not much antimatter in the Universe. 
Balloon born experiments detecting positrons have found since the 1980's a possible excess over secondary cosmic ray fluxes, the so called ``HEAT excess",
  which already in 1998 could be explained with WIMP annihilations, with WIMP masses above 200 GeV and annihilation  rate multiplied by  a ``boost factor"  $B>$ 30. PAMELA, a satellite carrying a magnetic spectrometer launched in 2006, 
reported  in 2008 an excess in the positron fraction $e^+/(e^+ + e^-)$  in the 10 to 100 GeV energy range~\cite{PAMELA-e} compatible with the ``HEAT excess". Shortly after,
 the ATIC collaboration announced a 6$\sigma$ excess in the 300-800 GeV range in 
the (e$^+ +$e$^-$) flux, with a sharp cutoff  at high energies (compatible with indicating the mass of annihilating WIMPs)  
which was later rejected by HESS and by Fermi. Fermi  measured the (e$^+ +$e$^-$) spectrum  from 20 GeV to 1 TeV with better accuracy than all preceding experiments and found a  excess of its own~\cite{FERMI} (without a sharp cutoff).  The PAMELA data on the antiproton to proton ratio and antiproton flux~\cite{PAMELA-pbar} is instead compatible with secondary cosmic rays, so whatever produces positrons should not produce an excess of antiprotons.

More than 500 papers have already been written trying to explain the PAMELA data, either with astrophysical or particle sources. The e$^+$ and e$^-$ come from less than 1 kpc, so must be produced locally.
 Pulsars or other supernova remnants nearby can account for the data~\cite{pulsars}.
It has also been suggested that secondary cosmic rays, such as e$^+$, could be accelerated at the sources of primary cosmic rays (leading to an enhancement of all secondary over primary ratios)~\cite{blasi}, an idea that will be confirmed or rejected soon (measuring ratios such as B/C). If the source is annihilating DM particles, only very tuned models survive all constraints. A simultaneous fit to the PAMELA, Fermi and HESS data requires the DM  to have mass of  TeV order,  to annihilate mostly into leptons of the 2nd or 3rd generation, $\tau^+\tau^-$ or 4$\mu$ or 4$\tau$ (not  into
 e$^+$e$^-$  or W pairs, because this would generate a sharp edge in the spectrum)~\cite{Meade}.  Thus, the DM must be ``leptophilic" either because the DM carries lepton number, or because of kinematics.  Moreover the annihilation rate must be larger than expected for thermal WIMPs by a boost  factor $B\simeq$10 to 10$^3$.  
 
 Astrophysical enhancements, due to nearby regions of larger DM density in the halo of our galaxy, cannot be larger than a few ($<10$), which is not enough. No boost factor at all is needed if WIMPs have a large annihilation cross section, both in the early Universe and in the dark galactic halo near Earth, which would produce a too small relic abundance for thermal WIMPs but could be fine if the pre-BBN cosmology is non-standard~\cite{kane}. Another possibility is that the annihilation cross section is larger in the dark halo at present, when WIMPs are more non-relativistic than in the early Universe, but it had the value necessary for thermal WIMPs to get the right relic density at the moment of freeze-out. This could be achieved if the DM annihilation cross section has a narrow resonance just below threshold, which is sampled more by low velocity particles than high velocity ones (see e.g.~\cite{ibe}) or by a ``Sommerfeld enhancement" of the cross section. The latter is due to the modification of the wave function of low velocity particles due to attractive long distance forces.  A classical analogy is that of particles approaching with speed $v$ a star of radius $R$ in the presence of gravity. Since the particles are deflected towards the star, the cross section $\sigma= \sigma_0 (1+ v^2_{escape}/ v^2)$ is larger than the geometrical cross section $\sigma_0= \pi R^2$, and for smaller velocities  $v << v_{escape}$ the enhacement is larger~\cite{Arkani-Hamed}. This mechanism works 
 for heavy almost degenerate neutralino-chargino  interactions, for which an attractive Yukawa force comes from  multiple t-channel W and Z exchange~\cite{Hisano}, or in the case the Yukawa force is due to the exchange of a light gauge or scalar boson.
 Besides all this requirements on the DM model, constraints imposed by the annihilation of the DM in the center of our galaxy  are only compatible with galactic halo models that predict a relatively small amount of DM in that region (core models as opposed to cusped ones)~\cite{Meade}. I should mention that also decaying DM has been considered (see e.g.\cite{Meade}). It  must decay mostly into leptons of the 2nd or 3rd generation, have multi-TeV mass and a very long lifetime $\tau \sim 10^{26}$ s.

Let us now present  ``A Theory of DM"~\cite{Arkani-Hamed}. In this model WIMPs with mass 500 to 800 GeV have exited states  with mass differences between 0.1 to 1 MeV and are  charged under a broken hidden gauge symmetry $G_{dark}$ with  gauge bosons $\phi$ (``dark photons") lighter than 1GeV. The bosons $\phi$ mediate new attractive forces which produce the Sommerfeld enhancement and are ``leptophilic"  because they are so light that can only decay into  $e^+ e^-$, $\mu^+ \mu^-$ or pions. This is a model made to explain simultaneously the DAMA annual modulation signal with ``inelastic" DM (IDM), the INTEGRAL data with ``eXciting'' DM (XDM) and the PAMELA positron fraction excess,  because the DM annihilates into a pair of $\phi$, each of which  decays afterwords, producing as final annihilation product mostly two   $\mu^+ \mu^-$ pairs (or pions).  Besides, the model provides signatures for the LHC (which depend on the particular realization of the model) such as GeV-dark Higgses and gauge bosons  which decay into visible particles and leptons, and also the excited WIMPs that decay producing many lepton jets with GeV invariant masses~\cite{Arkani-2}).

\section{Summary and outlook}

DM searches are independent and complementary to collider searches in multiple ways and they
are advancing fast. Lots of data lead to many hints, and the
data driven recent burst of model building has been due to the  difficulty in accommodating all recent hints.
So far, no firm DM signature has been  found but the many new models have opened our imagination and expectations
for things to come. The physics of DM and the physics needed at the electroweak scale may be different.
In any event, in most scenarios one can think of the LHC should find at least a hint of the new physics.
Whatever the LHC finds will  lead to a set of possible DM candidates and reject others.
Besides, DM may have several components to be found in different ways.
All possibilities are still open, hopefully not for long.

\end{document}